\documentclass[sigconf]{acmart}
\AtBeginDocument{%
  }

\setcopyright{acmlicensed}
\copyrightyear{2024}
\acmYear{2024}
\acmDOI{XXXXXXX.XXXXXXX}

\acmConference[RecSys 2024]{18th ACM Conference on Recommender Systems}{October 14–18,
  2024}{Bari, Italy}
\acmISBN{978-1-4503-XXXX-X/18/06}

\begin{document}

\title{TutorLLM: Customizing Learning Recommendations with Knowledge Tracing and Retrieval-Augmented Generation}

\author{Zhaoxing LI}
\email{zhaoxing.li@soton.ac.uk}
\orcid{0000-0003-3560-3461}
\affiliation{%
  \institution{University of Southampton}
  \streetaddress{University Road}
  \city{Southampton}
  \country{UK}
  \postcode{So17 1BJ}
}

\author{Vahid Yazdanpanah}
\affiliation{%
  \institution{University of Southampton}
  \city{Southampton}
  \country{UK}}
\email{v.yazdanpanah@soton.ac.uk}

\author{Jindi Wang}
\affiliation{%
  \institution{Durham University}
  \city{Durham}
  \country{UK}}
\email{jindi.wang@durham.ac.uk}

\author{Wen Gu}
\affiliation{%
  \institution{Japan Advanced Institute of Science and Technology}
  \city{Nomi}
  \country{Japan}}
\email{wgu@jaist.ac.jp}

\author{Lei Shi}
\affiliation{%
  \institution{Newcastle University}
  \city{Newcastle upon Tyne}
  \country{UK}}
\email{lei.shi@newcastle.ac.uk}

\author{Alexandra I. Cristea}
\affiliation{%
  \institution{Durham University}
  \city{Durham}
  \country{UK}}
\email{alexandra.i.cristea@durham.ac.uk}

\author{Sarah Kiden}
\affiliation{%
    \institution{University of Southampton}
    \city{Southampton}
    \country{UK}}
\email{sk3r24@soton.ac.uk}

\author{Sebastian Stein}
\affiliation{%
  \institution{University of Southampton}
  \city{Southampton}
  \country{UK}}
\email{ss2@ecs.soton.ac.uk}

\renewcommand{\shortauthors}{Zhaoxing Li et al.}

\begin{abstract}
The integration of AI in education offers significant potential to enhance learning efficiency. Large Language Models (LLMs), such as ChatGPT, Gemini, and Llama, allow students to query a wide range of topics, providing unprecedented flexibility. However, LLMs face challenges, such as handling varying content relevance and lack of personalization. To address these challenges, we propose TutorLLM, a personalized learning recommender LLM system based on Knowledge Tracing (KT) and Retrieval-Augmented Generation (RAG). The novelty of TutorLLM lies in its unique combination of KT and RAG techniques with LLMs, which enables dynamic retrieval of context-specific knowledge and provides personalized learning recommendations based on the student's personal learning state. Specifically, this integration allows TutorLLM to tailor responses based on individual learning states predicted by the Multi-Features with Latent Relations BERT-based KT (MLFBK) model and to enhance response accuracy with a Scraper model. The evaluation includes user assessment questionnaires and performance metrics, demonstrating a 10\% improvement in user satisfaction and a 5\% increase in quiz scores compared to using general LLMs alone.
\end{abstract}




\maketitle

\section{Introduction}
AI techniques are increasingly affecting various aspects of daily life, notably in educational environments. AI offers significant opportunities to enhance both the learning process and efficiency for students. Prominent among these AI applications are Large Language Models (LLMs), such as ChatGPT\footnote{https://chatgpt.com/}, Gemini \footnote{https://gemini.google.com/app
}, and Llama \footnote{https://llama.meta.com/}, which allow students to query a wide array of topics, thus offering unprecedented flexibility in learning. Unlike traditional search engines, LLMs enable students to ask nuanced and complex questions, engage in conversational interactions, and seek clarifications through follow-up questions, enhancing the depth and effectiveness of the learning experience. Despite these advantages, current LLMs face several challenges. These include generating inaccurate information (commonly referred to as ``hallucinations''), lack of personalization, and varying content relevance \cite{azamfirei2023large,floridi2020gpt, yan2024practical}. Specifically, these models often struggle with problems requiring high-level logical and mathematical problems, such as solving complex equations or providing step-by-step logical reasoning, and fail to tailor responses to individual learning levels, sometimes providing overly generalized answers or requiring extensive prompting to yield useful information \cite{yan2024practical}.

In contrast, traditional educational recommender systems utilize technologies such as Knowledge Tracing (KT) to track the learning trajectory of students and offer personalized recommendations. KT methods leverage historical interaction data to predict future learning actions \cite{corbett1994knowledge}. Despite their utility, these systems often fall short in terms of linguistic versatility and adaptability when compared to LLMs. Their responses are typically confined to a pre-defined set within their databases, limiting their ability to respond dynamically to a wide range of queries \cite{yan2024practical}. This limitation restricts the system's ability to adapt to individual learning needs in real-time, thereby reducing the effectiveness and engagement of the educational experience.

\begin{figure*}[h]
  \centering
  \includegraphics[width=\linewidth]{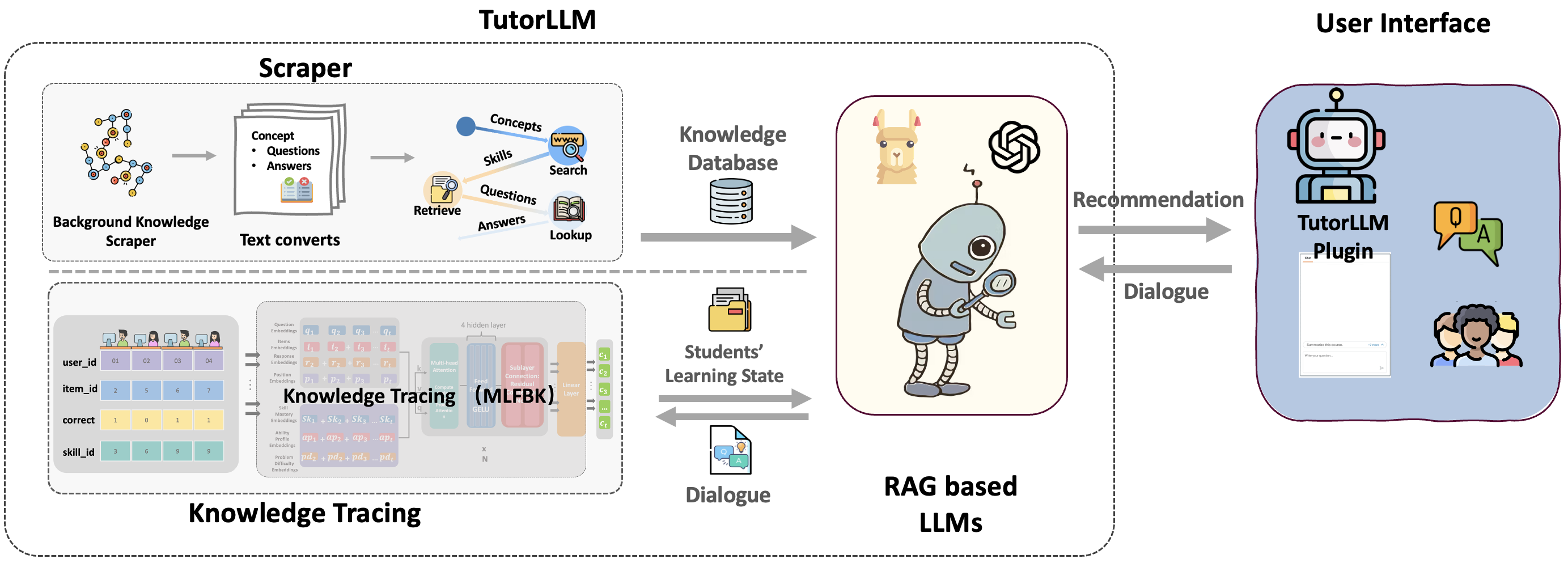}
  \caption{Overall architecture of TutorLLM.}
  \Description{the overall model architecture}
  \label{architecture}
\end{figure*}

To bridge the gap between the adaptability of LLMs and the personalized approach of educational recommender systems, we propose a novel framework: \textit{\textbf{the Personalized Educational Recommender LLM System (TutorLLM), based on Knowledge Tracing (KT) and Retrieval-Augmented Generation (RAG)}}. To the best of our knowledge, we are the first to incorporate KT technology into an LLM to achieve a personalized recommendation learning framework. The TutorLLM comprises three integral components. The first component is a KT model, which is developed based on the Multi-Features with Latent Relations BERT Knowledge Tracing (MLFBK) model \cite{li2023broader}. This component not only gathers data on student interactions and performance but also assimilates information from dialogues with LLMs, offering insights into student capabilities, learning states, and the complexities of the knowledge being acquired. The second is a Scraper. This component collects text content during online learning sessions to provide context-specific background knowledge that enhances the relevance and accuracy of the LLM's responses. The third component is an RAG Enhanced LLM. This component, utilizing the GPT-4 API, integrates inputs from both the Scraper and the KT module to deliver precise, personalized responses and learning content recommendations. Additionally, for even more tailored interaction, students can manually upload learning materials. Figure \ref{architecture} shows the overall model of the architecture. The text content provided by Scraper mitigates hallucinations by dynamically retrieving and incorporating context-specific knowledge from relevant course materials, ensuring accurate and reliable information. Additionally, the integration of knowledge tracing allows TutorLLM to deliver highly personalized recommendations and responses, tailored to each student's learning progress and needs.

To implement and evaluate this integrated approach, we developed a Chrome browser plugin that serves as an interface for students to interact directly with TutorLLM. Students could engage with TutorLLM by asking questions during online learning sessions. After the course, TutorLLM will provide students with personalized study material recommendations. Students can then decide whether to pursue further study based on these recommendations. Our evaluation, which involved a two-week field study with 30 undergraduate students in an online linear algebra course, used a crossover design to compare the effectiveness and user satisfaction between the general LLM approach and TutorLLM. Results showed a 10\% increase in user satisfaction for TutorLLM users compared to the general LLM approach, as measured by the System Usability Scale (SUS), and a 5\% improvement in academic performance, based on quiz scores.



\section{Related Work}

\textit{\textbf{Large Language Models (LLMs)}}, which have advanced natural language processing (NLP) and understanding (NLU), have significantly impacted various fields, including education \cite{kung2023performance}. These models, trained on vast text data, can generate human-like text, comprehend complex queries, and provide detailed explanations \cite{chang2024survey}. Examples like OpenAI's GPT-3 and GPT-4 are used in educational tools for tutoring, answering questions, and generating study materials \cite{brown2020language,chen2021survey}. However, LLMs can produce incorrect or misleading information (``hallucinations''), often lack personalized responses tailored to individual users' learning levels, and sometimes require significant prompting to be useful \cite{vaswani2017attention, yan2024practical}.

\textit{\textbf{Educational Recommender Systems}} provide personalized learning experiences by tailoring educational content based on students' needs, preferences, and progress \cite{ricci2011introduction}. Traditional systems use collaborative filtering, content-based filtering, or hybrid approaches to suggest relevant resources. Combining these systems with AI technologies like LLMs and KT models can significantly enhance personalization and performance in learning outcomes \cite{chen2020big, zhang2019deep,wang2024comparative,wang2023exploring,wang2024impact,wangchi2024}.

\textit{\textbf{Knowledge Tracing}} monitors and predicts students' knowledge states over time by tracking interactions with learning materials and assessments \cite{corbett1994knowledge, piech2015deep}. It enables personalized recommendations and targeted interventions to address the knowledge gaps of the students. Traditional methods like Bayesian Knowledge Tracing (BKT) \cite{pelanek2017bayesian} and Deep Knowledge Tracing (DKT) \cite{piech2015deep} have been widely used in Educational Recommender Systems, leveraging historical data to predict future performance and learning needs \cite{liu2021survey, pandey2019machine}. Recent models, such as Multi-Features with Latent Relations BERT Knowledge Tracing (MLFBK), enhance the accuracy and depth of predictions \cite{li2021survey,chen2023improving,lee2022monacobert,tiana2021bekt,li2023broader,li2023deep,li2023towards,li2024lbkt,li2023sim,li2024integrating}.

\section{Implementation of TutorLLM}
TutorLLM consists of three main components: the Scraper Model for collecting educational content, the KT for predicting students' learning states, and the RAG based LLM for dynamically retrieving information and tailoring personalized responses.

\begin{figure}[h]
  \centering
  \includegraphics[width=0.7\linewidth]{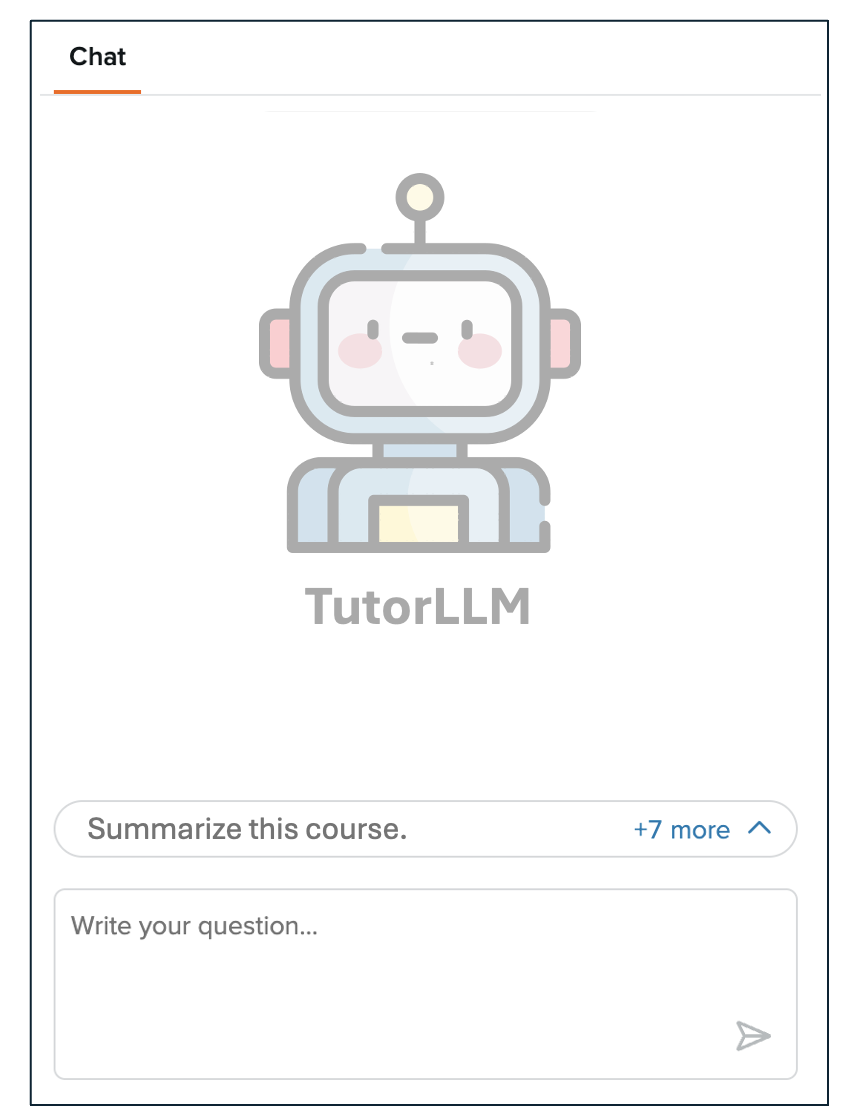}
  \caption{User Interface of the TutorLLM.}
  \Description{the UI of the TutorLLM}
  \label{ui}
\end{figure}

\textit{\textbf{Overall Methodology.}}
The motivation of our approach is to enable LLMs to comprehend the student's learning state (or knowledge master state) and the specific content of the ongoing course, thereby furnishing contextualized responses and tailored learning content recommendations. Thus, our method is structured into three distinct components. Firstly, the KT component utilizes an algorithm to trace students' learning state. This encompasses tracing skill mastery, ability profiles, problem difficulty, and predicting the next most probable action for the student. We employ MLFBK as our KT method within this component. Secondly, the Scraper function is designed to gather and organize the text information from the online course platform, including captions and subtitles of videos embedded in the web pages, into a background knowledge base. After receiving student action sequence data from the KT component, the TutorLLM first synthesizes the student's current learning status, focusing on identifying weak knowledge areas and predicting the student's potential next actions. Finally, the model gives personalized answers and recommendations based on the background knowledge base and students' learning state. Additionally, the model recommends additional learning materials to students upon request or after each study session. Based on the above ideas, we built a Chrome browser plug-in. When students open an online course website, they can open our TutorLLM by clicking the button on the right-hand side of the address bar. Figure \ref{ui} shows the interface of the TutorLLM. Detailed insights into the functionalities of these three components are provided below:

\textit{\textbf{Scraper Model}}:
The Scraper component of TutorLLM was developed to autonomously collect and organize textual content from online course web pages. It extracts text information, including crucial captions and subtitles, to build a rich background knowledge base. This functionality is built on the Reader API from Jina AI\footnote{https://github.com/jina-ai/reader
}. In operation, the Scraper dynamically interacts with educational websites as a Chrome plugin activated by students. Upon visiting a relevant page, the Scraper processes the content, converting Uniform Resource Locators (URLs) to LLM-friendly inputs that include structured text and contextual captions. This ensures that every piece of extracted content is optimized for use by the large language model, providing accurate, up-to-date information that reflects the current scope of the course materials.

\textit{\textbf{Knowledge Tracing Model}}: For the knowledge tracing part, we use MLFBK\cite{li2023broader} to incorporate multi-feature embedding and latent Relations to capture students' learning state. MLFBK consists of three parts: embedding, BERT-based architecture, and correctness sequence output. 

Within the embedding part, there are two components: Multi-Features embedding and Latent-Relations embedding. In the multi-feature embedding component, four distinct features, including student\_id, skill\_id, question\_id and response\_id are integrated. The latent relations embedding component employs a feature engineering method to extract three significant relations: skill mastery, ability profile, and problem difficulty. Skill mastery is modeled based on Bayesian Knowledge Tracing, while the ability profile is encoded using past performance in various time intervals and updated via the K-means algorithm \cite{steinley2006k}. Problem difficulty is quantified on a scale of 1 to 10 (where 1 is easy and 10 is extremely difficult), derived from success rates. Additionally, the overall model integration encompasses a combination of embeddings such as question, item, response, skill mastery, ability profile, and problem difficulty embeddings, culminating in creating a final input embedding. 

In the BERT-based architecture, encoder blocks leverage a pre-LN Transformer architecture \cite{vaswani2017attention}, incorporating monotonic convolutional multi-head attention followed by fully connected layers with LeakyReLU activation. Monotonic multi-head attention, in conjunction with mixed attention and monotonic attention, is utilized for sequence data representation.

In the correctness sequence output part, we initially acquire the prediction regarding the student's sequence action data. Subsequently, the prediction action will append to the end of the student's historical action sequence, generating the complete output sequence.

\textit{\textbf{RAG Large Language Model:}}
The large language model component, powered by the GPT-4 API, utilizes a Retrieval-Augmented Generation (RAG) \cite{lewis2020retrieval} approach to produce responses.

Upon activation, the model receives a processed input from the knowledge tracing component, which includes detailed insights about the student’s current mastery of various skills and predicted next actions. The retrieval process begins with the model querying the background knowledge base, which has been populated by the Scraper model with data from online course materials. This ensures that the responses are not only based on generic information but are enriched with up-to-date, course-specific content that enhances learning effectiveness.

The LLM then synthesizes the information from the knowledge base with the student's specific learning context to generate responses. These responses could range from direct answers to queries, explanations of complex topics, or hints to guide problem-solving. Furthermore, the model actively offers study recommendations based on the student’s learning progress and areas of difficulty, which could include suggestions for revisiting certain topics or advancing to new content based on the student’s readiness.

\section{User Study}

\textit{\textbf{Participants \& Study Setup.}} To evaluate the effectiveness of the TutorLLM educational tool on student learning outcomes, 30 first-year undergraduate students from the University of Southampton enrolled in a linear algebra module were randomly allocated to one of three groups in a controlled experiment. The control group exclusively used general LLMs (the general web version of chatGPT4) for the duration of the study. Experimental Group 1 was exposed to general LLMs during the first week, followed by TutorLLM in the second week, whereas Experimental Group 2 utilized TutorLLM throughout both weeks. The key objective was to determine if integrating TutorLLM could enhance learning performance compared to the general LLMs. 


\textit{\textbf{Student Performance.}} Over a two-week period, students explore a new segment of linear algebra each day. They utilize our TutorLLM to address their queries daily during their studies. At the course's conclusion, TutorLLM suggests pertinent educational resources and exercises for further learning. Students have the autonomy to decide whether to engage with these additional materials, unlike those in the control group who use a general LLM model and do not receive such recommendations. Each day culminates with a quiz testing the knowledge acquired, leading up to a comprehensive exam covering all topics at the end of the two weeks. The assessment framework includes 15 tests overall, with each daily test comprising 10 questions and the potential to score up to 100 points. These daily quizzes and the final comprehensive test together determine the students' overall performance.

\textit{\textbf{User Study.}} A questionnaire was administered to students in Group 1 and Group 2 to evaluate the effectiveness of the TutorLLM and its impact on student satisfaction. The survey was designed to gather quantitative and qualitative data on students' experiences with the TutorLLM. It incorporated a System Usability Scale (SUS) \cite{bangor2008empirical} and a User Experience (UX) \cite{vermeeren2010user} questionnaire to provide a comprehensive assessment. SUS, a validated tool, measures user perceptions of usability, focusing on user-friendliness and comprehensibility, with higher SUS scores indicating a more intuitive interface and enhanced user control. UX was measured from three perspectives, which are User Satisfaction (US), Comfort Level (CL) and Continue Willingness (CW). US was evaluated using a 5-point Likert scale to assess the overall satisfaction with the tool, where a positive US score reflects the assistant’s ability to meet or exceed user expectations regarding usefulness and responsiveness. CL was rated on a 5-point scale to gauge user comfort during initial textual interactions on online platforms, providing critical insights for optimizing chat interfaces and enhancing interaction experiences \cite{bergin2005influence}. CW, measured on a 5-point scale, captures users’ willingness to continue interactions after the initial conversation \cite{freiermuth2006willingness}.

\section{Results}

\begin{figure}[h]
  \centering
  \includegraphics[width=\linewidth]{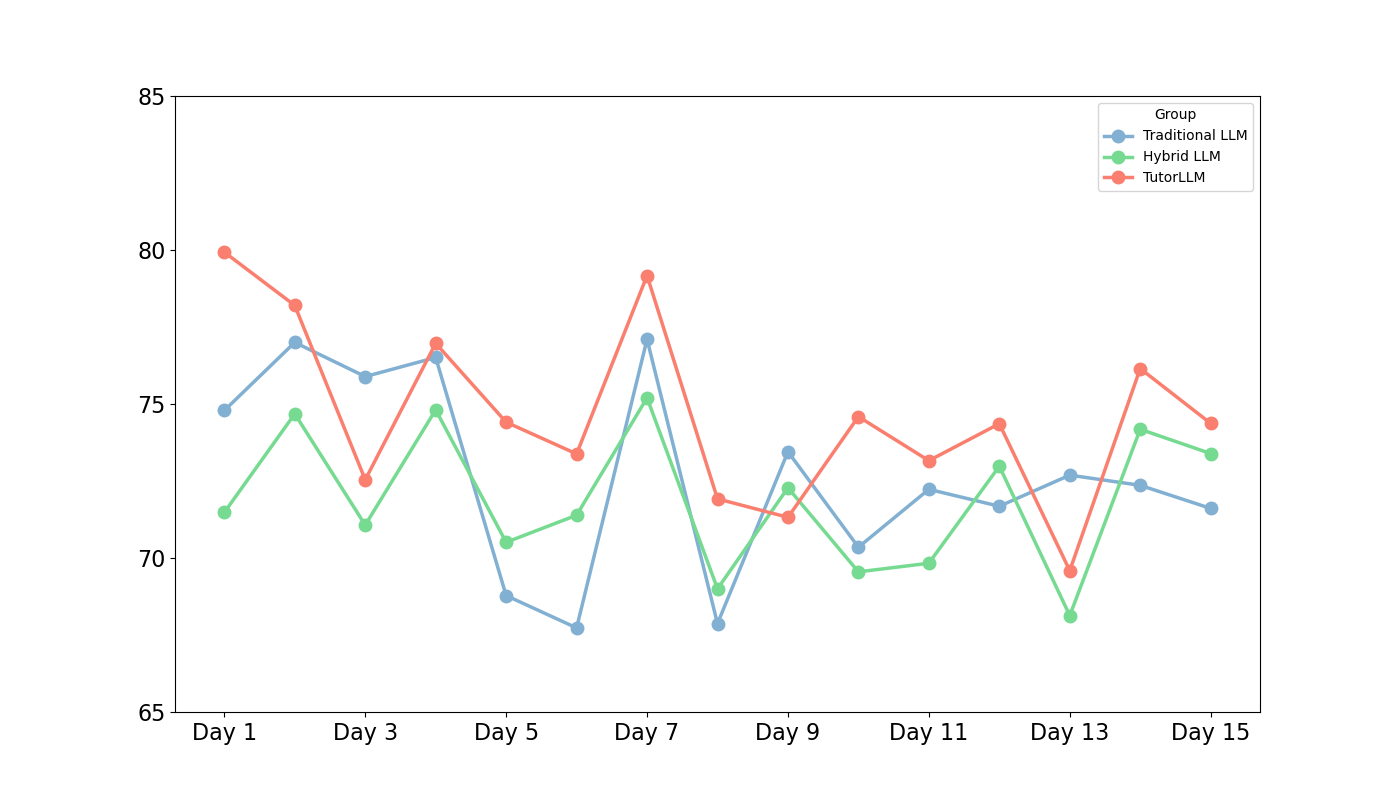}
  \caption{Daily Mean Scores of User Performance Across Different Study Groups}
  \Description{the UI if the TutorLLM}
  \label{Ave}
\end{figure}

\textit{\textbf{Test Results.}} We evaluated the student performance of students across three control groups through 15 tests, including a final examination. The groups comprised students utilizing general LLMs, a hybrid LLM approach, and TutorLLM throughout the study. Our initial analysis revealed that the TutorLLM group achieved the highest overall mean score of 74.48, followed by the hybrid LLM group at 72.81 and the general LLM group at 71.97. To determine the statistical significance of the differences observed in the final exam scores among the different groups, we conducted an analysis of variance (ANOVA). This analysis produced an F-statistic of 0.795 and a p-value of 0.462, indicating no significant variance in performance across the groups. Figure \ref{Ave} displays the daily mean scores for each study group. Although the differences were not statistically significant, there was a discernible trend toward improved performance in the TutorLLM group. These results suggest that additional research, potentially involving a larger sample size, different study designs, or a longer test duration, is necessary to definitively ascertain the effectiveness of integrating TutorLLM in enhancing academic performance.

\textit{\textbf{Usability and User Experience.}} From the usability perspective, the SUS results indicated a generally positive user experience, with an average SUS score of 76.35 and a median score of 79, reflecting high usability. The scores showed moderate variability, with a standard deviation of 14.46, and the majority of scores ranged between 61.89 and 90.81, demonstrating consistency in positive evaluations. The Shapiro-Wilk test confirmed the normality of the score distribution ($p = 0.721$), supporting the validity of the usability assessment. A bootstrap analysis estimated the median's 95\% confidence interval to be between 68.5 and 83, further validating the positive user feedback on the TutorLLM’s usability.

From the user experience perspective, the analysis of CL, CW, and US showed average scores of 3.50 for CL, 3.40 for CW, and 3.61 for US, with respective standard deviations of 1.00, 0.99, and 0.82. Correlation analysis revealed moderate positive correlations among the metrics CL and CW at 0.53, CL and US at 0.45, and CW and US at 0.40, suggesting that users felt more comfortable and satisfied with the app and were more willing to continue using it.

\section{Discussion}
While our study did not show statistically significant improvements in academic performance with TutorLLM, the increased engagement and higher average scores suggest its potential benefits. We monitored the duration of LLM usage among students. Users of TutorLLM spent 36\% more time engaging with the system compared to those using general LLMs. This increased engagement suggests that students were more satisfied with TutorLLM, making them more willing to invest additional time in the system. Additionally, the learning materials recommended by our TutorLLM appear to have a positive impact on students' learning performance to a certain extent. These findings underscore the importance of further refining AI-driven educational tools to better adapt to individual learning needs and contexts. It also could redefine personalized learning, making educational interactions more effective and engaging through tailored content and intelligent response systems.


\section{Conclusion}
In this paper, we introduced TutorLLM, a novel framework that integrates Large LLMs with KT and RAG to enhance personalized learning. Our main contributions include being the first to combine LLMs with KT to improve personalization. We demonstrated the practical application of TutorLLM through a Chrome browser plugin and validated its effectiveness in a two-week study with undergraduate students. Future research should focus on refining TutorLLM's personalization features, testing its effectiveness across various disciplines, and addressing challenges such as integrating existing technologies, ensuring data privacy, and providing educator training. 


\bibliographystyle{ACM-Reference-Format}
\bibliography{software}










\end{document}